\documentstyle[12pt,fleqn,graphicx]{article}

\newcommand*{\FL}{}
\newcommand*{\g}{\mbox{\slshape g}}
\newcommand*{\bmv}[1]{\mbox{\boldmath$#1$}}

\begin{document}
\vspace*{2cm}
\noindent
{\large\bf
Investigation of transfer coefficients for many-component
dense systems of neutral and charged hard spheres}
\vspace*{1.5ex}

\noindent A.E.Kobryn, M.V.Tokarchuk, Y.A.Humenyuk
\vspace*{1.5ex}

\noindent
Institute for Condensed Matter Physics of the National Academy of
Sciences of Ukraine 1~Svientsitskii St., UA-79011 Lviv, Ukraine
\vspace*{3ex}

Investigation of transfer processes in dense weakly or fully ionized
many-component gases is topical in view of construction and improvement
of new gaseous lasers, plasmo-chemical reactors for ozone synthesis, and
air cleaning of both nitrogen and carbon oxides and chlorine compounds.

A study and calculation of transfer coefficients taking into account the
interparticle interaction nature is one of the major problems in this
investigation. Recent papers by Murphy [1-4] are devoted to these
calculations for some special low-density mixtures basing on the
Boltzmann kinetic equation for point particles and using accurate
interparticle potentials of interaction. Nevertheless, there are some
problems in calculation of transfer coefficients for high-density
mixtures. They are caused mainly by the fact that the Boltzmann kinetic
equation is suitable for rarefied gases and plasmas only.

In present work a calculation of transfer coefficients for many-component
dense gases for charged and non-charged hard spheres is carried out using
the Enskog-Landau kinetic equation which takes into account realistic
particle sizes.

The Enskog-Landau kinetic equation was obtained in [5,6] for a
one-component system of charged hard spheres in the electron compensation
field starting from the Bogolubov hierarchy with modified boundary
conditions in the pair collision approximation. In paper [7] this
equation was generalized for the case of many-component system and
usually reads:
\FL
\begin{equation}
\left[\frac{\partial}{\partial t}+{\bmv v}_a\cdot
\frac{\partial}{\partial{\bmv r}_a}\right]f_a(x_a;t)=
\sum_{b=1}^M
\left[
I^{(0)}_{\rm E}(f_a,f_b)+I^{(1)}_{\rm
E}(f_a,f_b)+I_{\rm MF}(f_a,f_b)+I_{\rm L}(f_a,f_b)
\right],
\end{equation}
where $f_a(x_a;t)$ is the one-particle distribution function of
$a$-component ($x\equiv\{\bmv{r},\bmv{p}\}$), $M$ is a number of
components.

The collision integral in the right-hand part of (1) consists of several
terms. Such a structure is produced due to the additivity of the
interaction potential which is modelled for the mixture of charged hard
spheres as a sum of the short-range part (hard spheres) and the
long-range one (Coulomb particles). $I_{\rm E}^{(0)}$ and $I_{\rm
E}^{(1)}$ are the zeroth and first expansion terms of the Enskog
collision integral with respect to the spatial inhomogeneity [5]. $I_{\rm
MF}$ is the collision integral of the kinetic mean field theory KMFT
[8,9]. It is the first order on the long-range interaction. The last term
$I_{\rm L}$ is the generalized Landau collision integral and it is the
second order on the long-range interaction.

The kinetic equation (1) can be solved with operating the standard
Chapman-Enskog method [10]. The unknown distribution function in the
first approximation can be taken as
\FL
\begin{equation}
f^{(1)}_a(\bmv{r},\bmv{v}_a,t)=
f^{(0)}_a(\bmv{r},\bmv{v}_a,t)[1+\phi_a],
\end{equation}
where $f_a^{(0)}$ is the zeroth approximation. Usually, it is chosen as
the local-equilibrium Maxwell distribution function. $\phi_a$ is a
correction which is expressed through the Sonine-Laguerre polynomials
[10].

Having the solution to the kinetic equation (1), one can calculate the
stress tensor and the heat flow vector in a system. These quantities are
expressed via such transport coefficients as bulk $\kappa$ and shear
$\eta$ viscosities and thermal conductivity $\lambda$, respectively:
\FL
\begin{eqnarray}
\kappa&=&
\frac{4}{9}\sum_{a,b=1}^M
\sigma_{ab}^4\g_2^{ab}n_an_b\sqrt{2\pi\mu_{ab}k_{\rm B}T}=
\sum_{a,b=1}^M\kappa_{ab},\\
\eta&=&\frac{3}{5}\kappa+\left(\frac{1}{2}
\sum_{a=1}^Mn_aB_{a}(0)
+\frac{4\pi}{15}\sum_{a,b=1}^M
\sigma_{ab}^3\g_2^{ab}n_an_{b}
\frac{\mu_{ab}}{m_b}B_{b}(0)\right)k_{\rm B}T,\\
\lambda&=&\sum_{a,b=1}^M \frac{3k_{\rm B}\,\kappa_{ab}}{m_{a}+m_b}-
\sqrt{2k_{\rm B}^3T}\Bigg(\frac{5}{4}\sum_{a=1}^M\frac{n_a}{\sqrt m_a}
[A_{a}(1)-A_{a}(0)]+{}\\
&&\frac{2\pi}{3}\sum_{a,b=1}^M
\sigma_{ab}^3\g_2^{ab}\frac{n_an_b}{m_a+m_b}\Big[\frac{3\mu_{ab}}{\sqrt{m_b}}
A_b(1)-\sqrt{m_b}A_b(0)\Big]\Bigg),\nonumber
\end{eqnarray}
where $\sigma_{ab}=(\sigma_a+\sigma_b)/2$, $\sigma_a$, $\sigma_b$ -- hard
sphere diameters, $\g_2^{ab}$ -- quasiequilibrium binary correlation
function, $n_a$ -- particle number density, $\mu_{ab}=m_am_b/(m_a+m_b)$
-- reduced mass, $A_a(0)$, $A_a(1)$ and $B_a(0)$ are zeroth and first
coefficients of expansion of the correction $\phi_a$ on the
Sonine-Laguerre polynomials.

In numerical calculations of transfer coefficients (3)--(5) for some
binary and ternary mixtures of neutral and charged hard spheres we
considered their temperature and concentration ratio dependences. The
values of hard sphere diameters were fixed: $\sigma_{\rm He}=2.15$\AA,
$\sigma_{\rm Ar}=3.405$\AA, $\sigma_{\rm Kr}=3.67$\AA, $\sigma_{\rm
Xe}=3.924$\AA. These values were borrowed from [11].

Calculating $A_a(0)$, $A_a(1)$ and $B_a(0)$ we faced with the problem of
divergency of the so-called $\Omega$-integrals at large distances. To
avoid this circumstance we should change upper limit of integration and
use finite screening radius $D$ instead of infinity. This screening
radius, in contrast to the Debye formula for point-like particles,
explicitly takes into account realistic sizes of charged particles in
accordance with [12]. Contact values of the binary correlation function
in expressions (3)--(5) were borrowed from [13].

Not only the transport coefficients $\kappa$, $\eta$, and $\lambda$ were
studied. Numerical calculation for thermal diffusion $D^\alpha_{\rm T}$
and mutual diffusion coefficients $D^{\alpha\beta}$ has been performed as
well. For two- and three-component mixtures of neutral and charged hard
spheres we studied their dependences on density, temperature, and
concentration ratio of some mixture components [14]. Figs. 1 and 2 show
our calculations in brief.

\begin{figure}[!ht]
\begin{center}
\includegraphics*[bb=57 61 531 478,%
    width=0.45\hsize
    ]{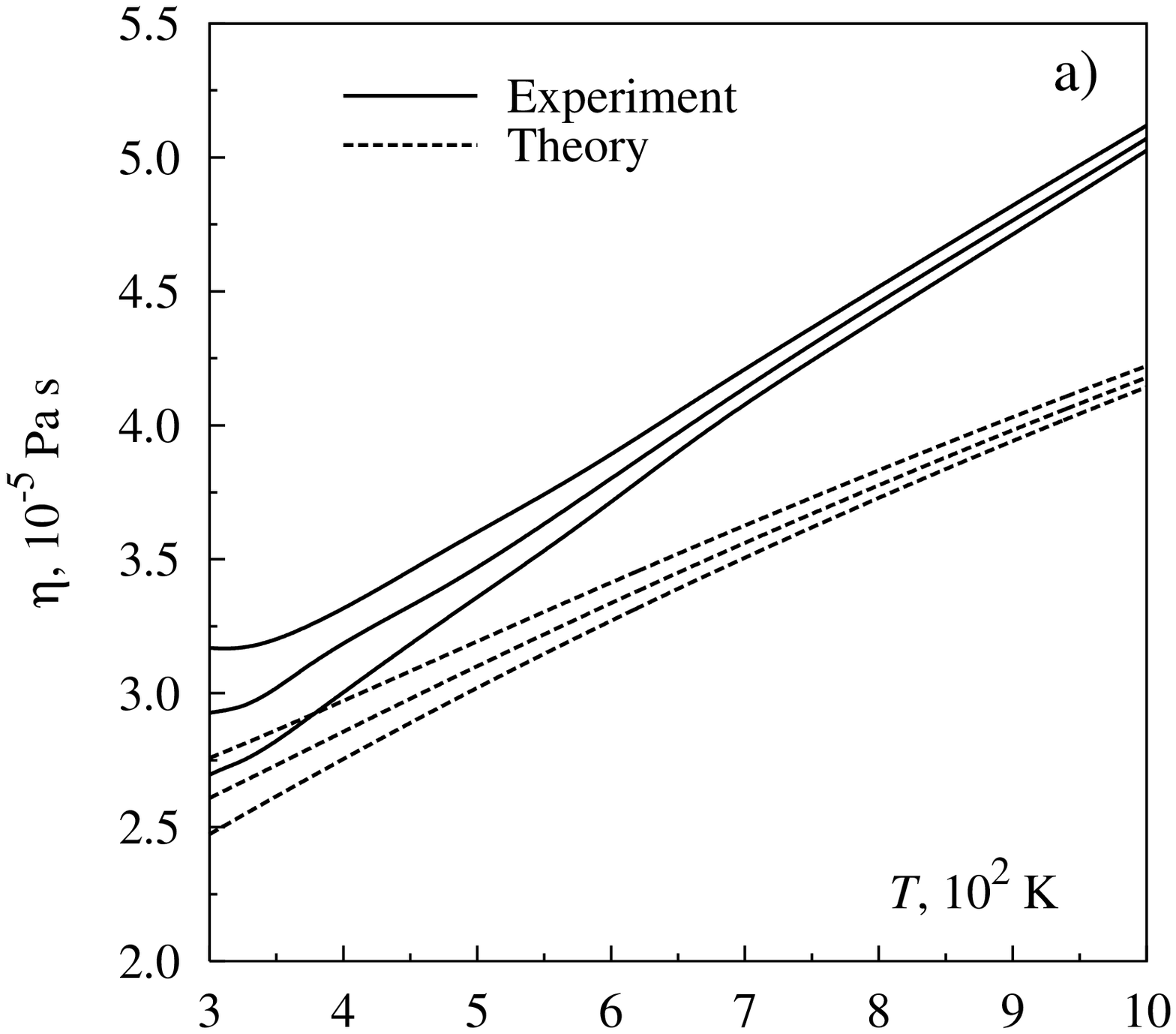}%
\hfill
    \includegraphics*[bb=57 61 531 478,%
    width=0.45\hsize,
    ]{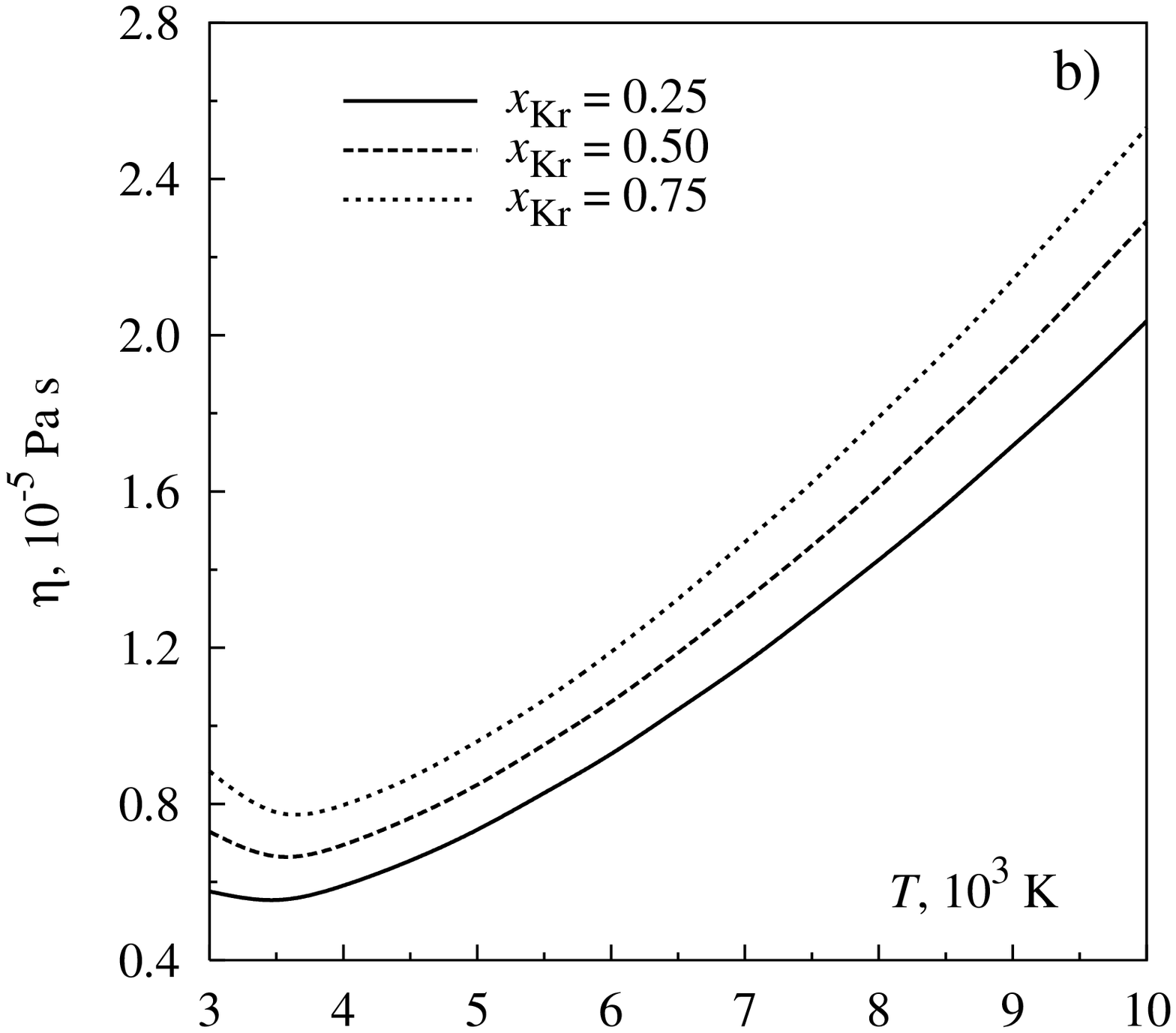}
\end{center}
\begin{center}
    \includegraphics*[bb=57 62 533 478,%
    width=0.45\hsize
    ]{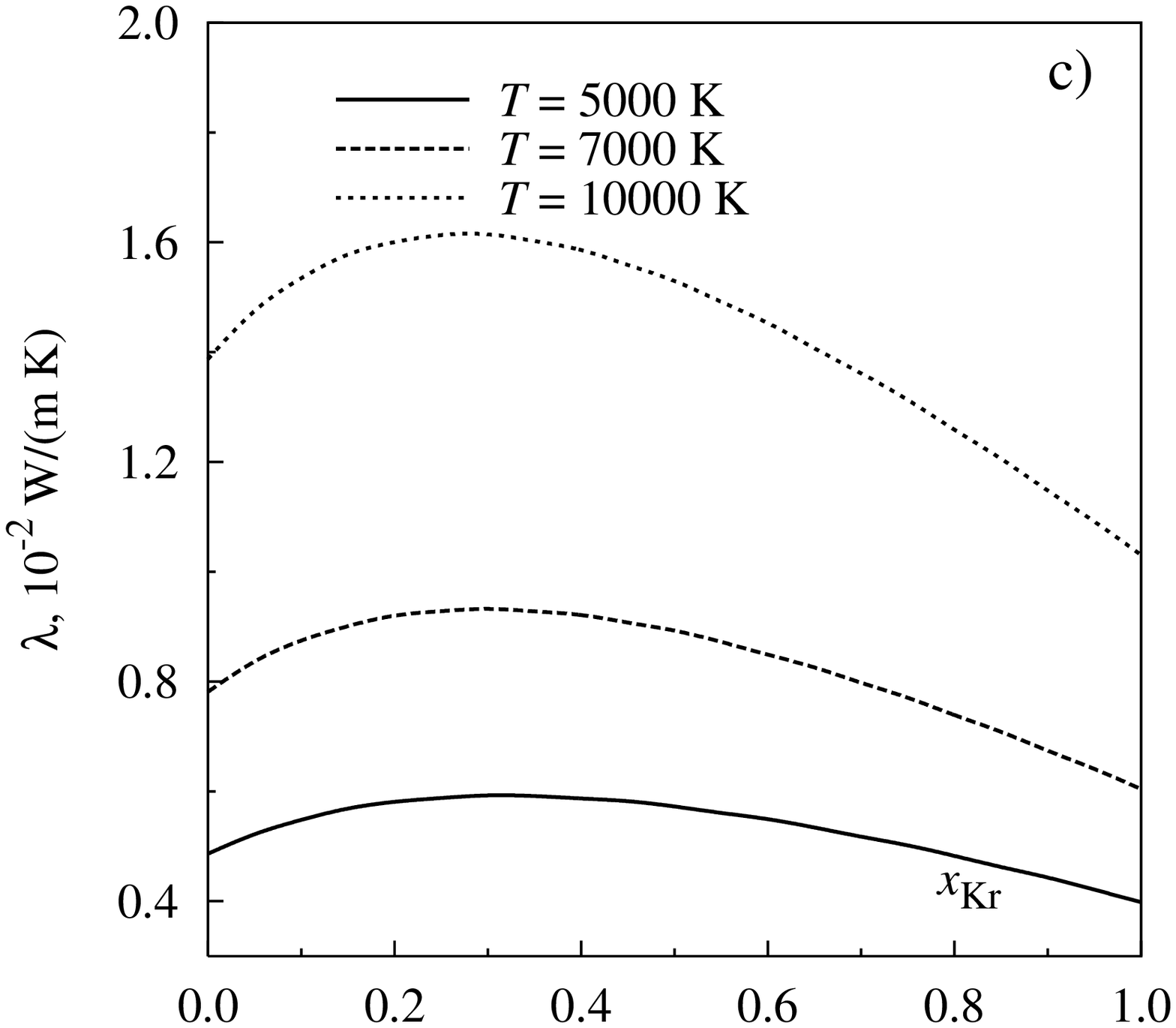}%
\hfill
    \includegraphics*[bb=57 61 531 478,%
    width=0.45\hsize
    ]{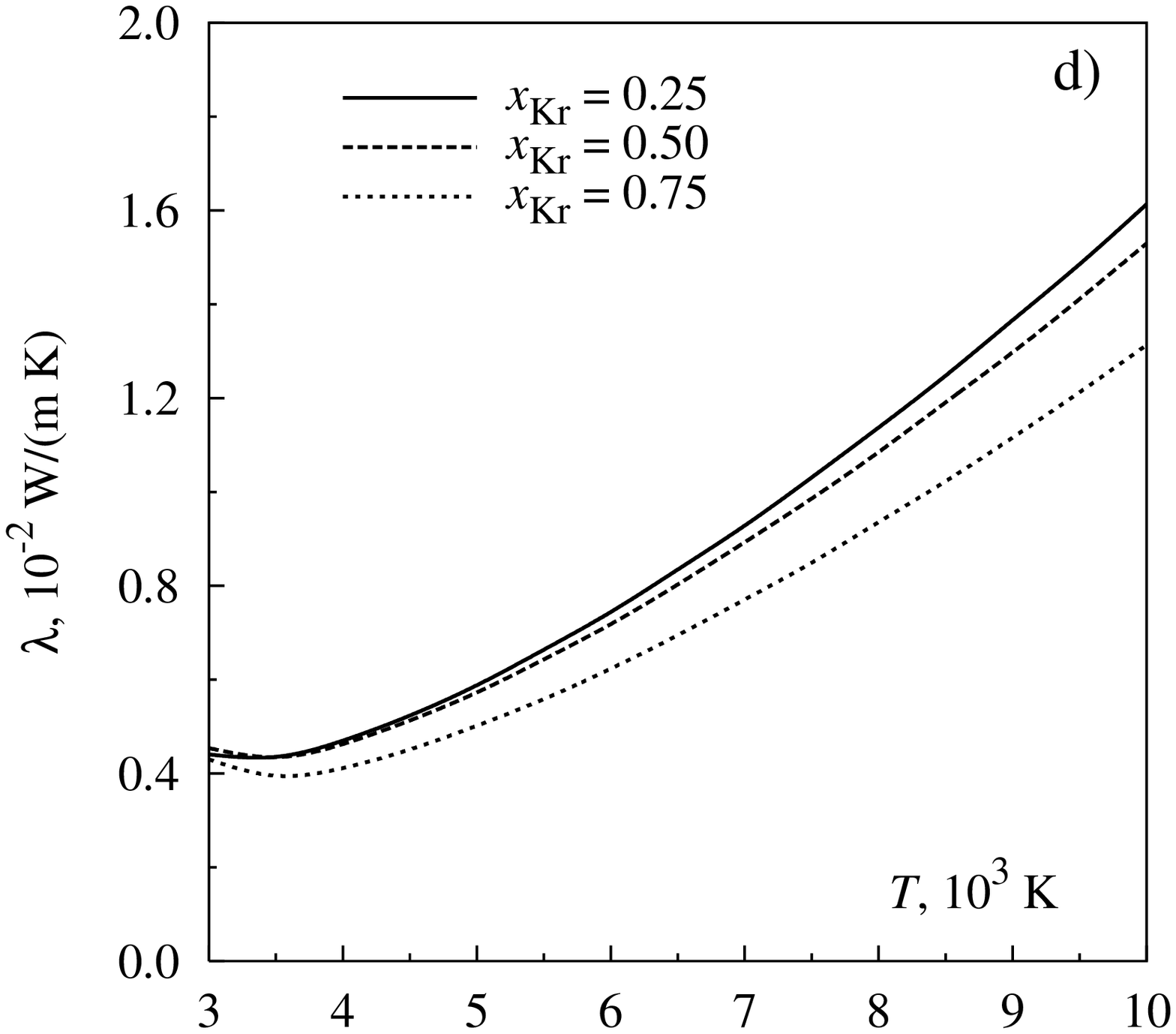}
\end{center}
\caption{\small a) -- temperature dependence of shear viscosity $\eta$
for mixture 40\%H$_2$, 60\%Ar at pressures $P=300$, 400, 500 atm.,
respectively. Curves are plotted in order of increasing pressure. Though
this model is not entirely suitable as a hard spheres model, because the
H$_2$-molecule is not spherical, nonetheless obtained results give an
information about the successfulness of the application of the model for
nonspherical molecules. Somewhat worse agreement is between the theory
and experimental data [15] when temperature increases. At high
temperatures the role of the attractive part of interparticle interaction
potential becomes weaken in comparison with the repulsive one. However,
this behaviour can be explained by ignoring in the theory the temperature
and concentration dependence of hard sphere diameters for molecules of Ar
and H$_2$. At the same time it is known that $\sigma$ depends both on $T$
and $n$. b), c) and d) -- temperature and concentration ratio dependences
of shear viscosity $\eta$ and thermal conductivity $\lambda$ coefficients
for the mixture Ar$^+$-Kr$^+$ at the fixed total density
$n=2\cdot10^{26}{\rm m}^{-3}$. These calculations were performed for
three fixed concentration ratios of the heavy component $x_{\rm
Kr^+}=0.25$; 0.5; 0.75, $x_{\rm Kr^+}=n_{\rm Kr^+}/(n_{\rm Ar^+}+n_{\rm
Kr^+})$. The coefficients $\eta$ and $\lambda$ have very similar
temperature dependence. But concentration ratio dependence for $\lambda$
has a different curve fit. More detailed analysis is given in [14].}
\end{figure}
\begin{figure}[!ht]
\begin{center}
    \includegraphics*[bb=57 61 546 422,%
    width=0.48\hsize
    ]{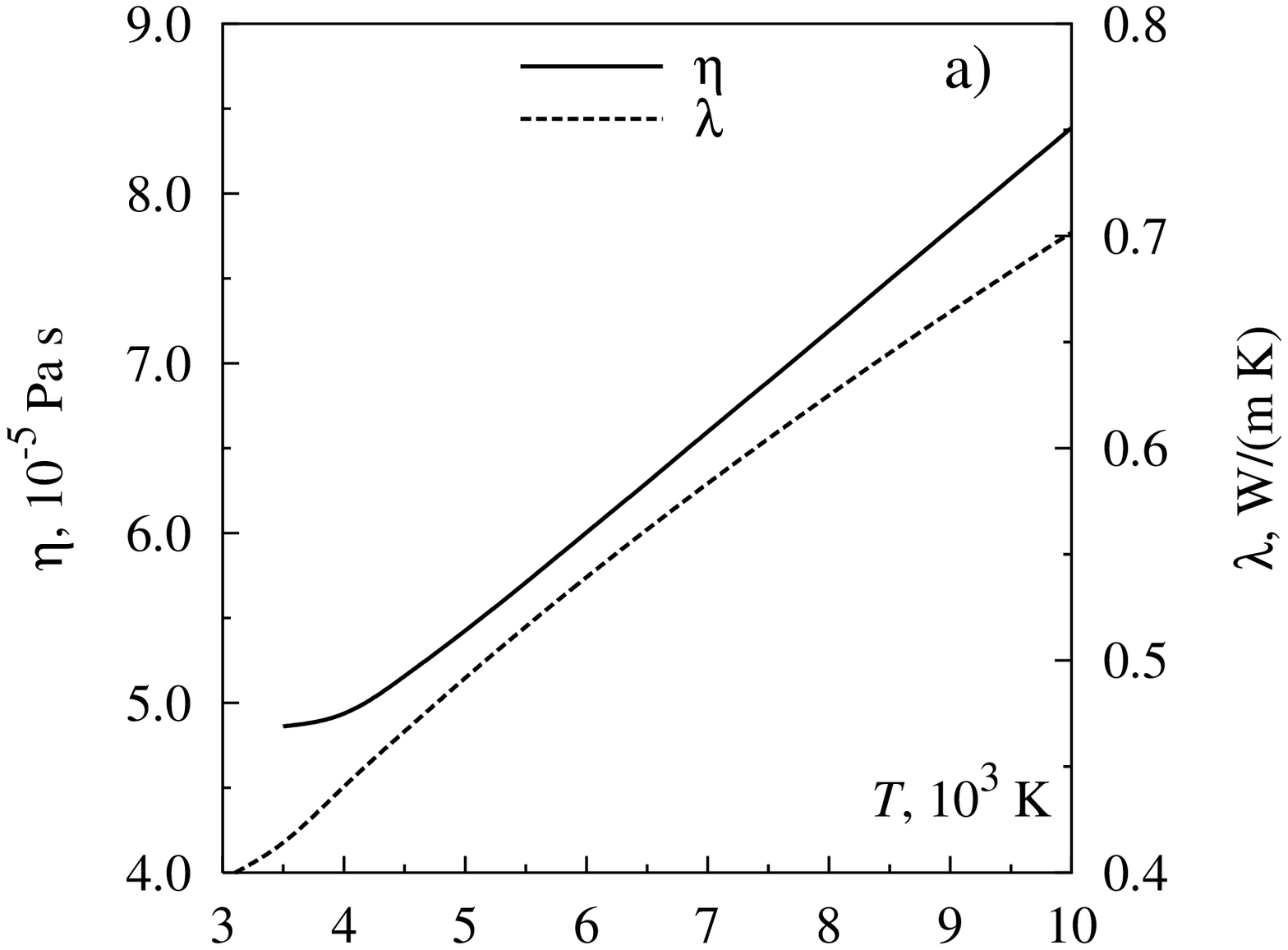}
\hfill
    \includegraphics*[bb=54 61 547 422,%
    width=0.48\hsize
    ]{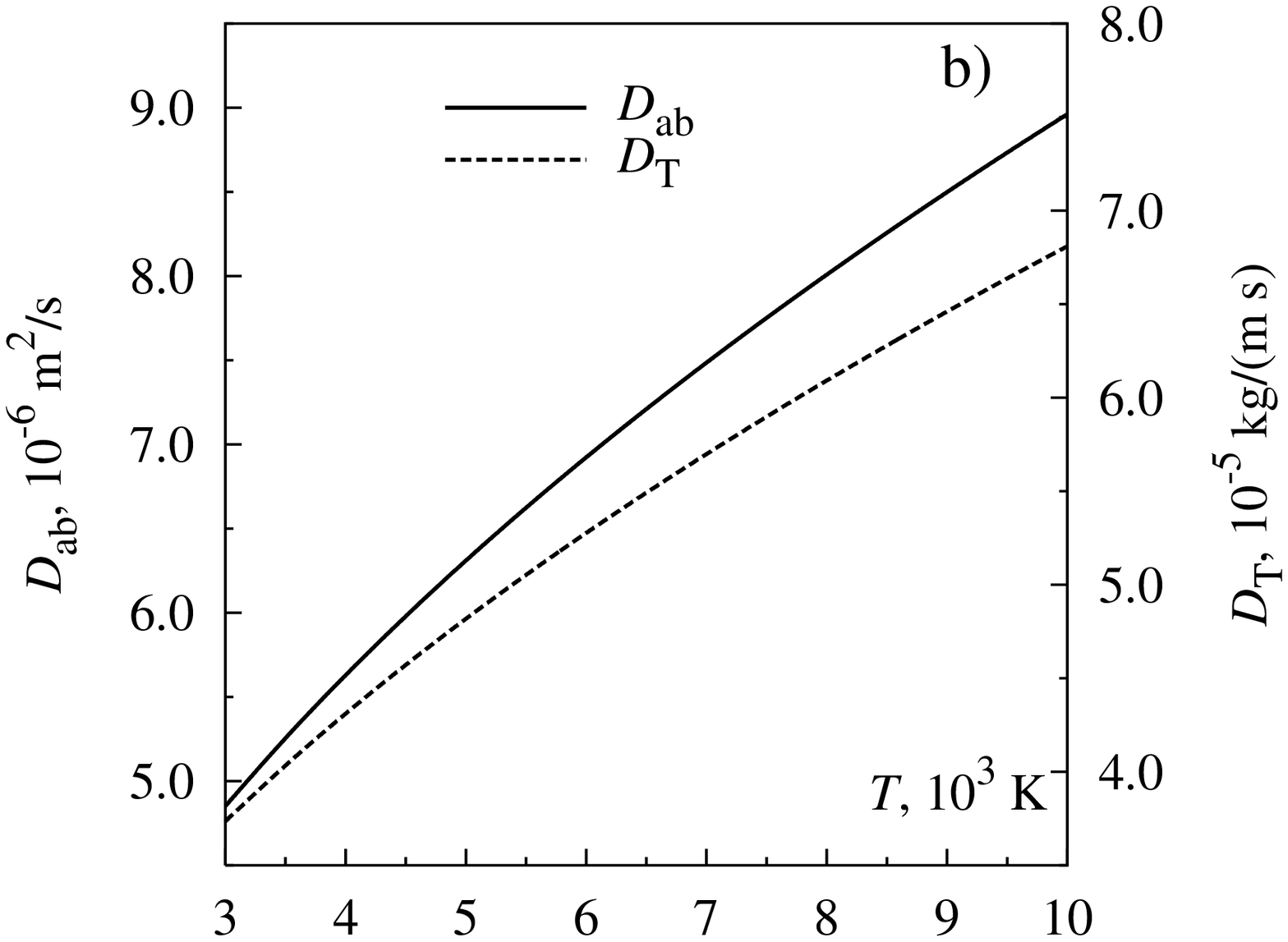}
\end{center}
\caption{Temperature dependences of transport coefficients: shear
viscosity $\eta$ and thermal conductivity $\lambda$ a), and mutual
diffusion coefficient $D_{\rm ab}$ and thermal diffusion coefficient
$D_{\rm T}$ b) for the system 26.7\%Ar$^+$, 73.3\%He at
$n=1.0487\cdot10^{27}$ m$^{-3}$, or $\Delta=0.00976$.}
\end{figure}

The obtained Enskog-Landau kinetic equation for charged hard spheres
turned out to be very useful for several purposes. First of all, the
collision integral of this equation does not contain a divergency at
small distances. Secondly, the normal solution and all transport
coefficients have analytical structure. They can be easily used to study
some specific systems. Finally, the analytical structure of transport
coefficients allows us to find fast and easily systems, which can be best
described by the obtained kinetic equation, as well as density and
temperature ranges, where the agreement between the theory and
experimental data is the closest.

A very important step in this theory is to calculate a dynamical
screening radius in a system. Partially this problem has been already
solved in our recent paper [16].


\begin{thebibliography}{00}
\itemsep-2pt
\footnotesize
\bibitem{1} A.B.Murphy, Phys. Rev. E, 48 (1993) 3594.
\bibitem{2} A.B.Murphy, C.J.Arundel, Plasma Chem. \& Plasma Processing,
        14 (1994) 451.
\bibitem{3} A.B.Murphy, Plasma Chem. \& Plasma Processing, 15 (1995) 279.
\bibitem{4} A.B.Murphy, IEEE Transactions on Plasma Sci., 25 (1997) 809.
\bibitem{5} D.N.Zubarev, V.G.Morozov, I.P.Omelyan, M.V.To\-kar\-chuk,
    Teor. Mat. Fiz., {87} (1991) 113.
\bibitem{6} A.E.Kobryn, V.G.Morozov, I.P.Omelyan, M.V.Tokarchuk, Physica A,
        230 (1996) 189.
\bibitem{7} A.E.Kobryn, I.P.Omelyan, M.V.Tokarchuk, Physica A, 268 (1999)
        607.
\bibitem{8} J.Karkheck, G.Stell, J. Chem. Phys., 75 (1981) 1475.
\bibitem{9} G.Stell, J.Karkheck, H. van Beijeren,
        J. Chem. Phys., 79 (1983) 3166.
\bibitem{10} J.H.Ferziger, H.G.Kaper, Mathematical theory of transport
        processes in gases. (North Holland, Amsterdam, 1972).
\bibitem{11} M.F.Pas, B.J.Zwolinski, Mol. Phys., 73 (1991) 471.
\bibitem{12} L.Blum, J.S.H\o ye, J. Phys. Chem., 81 (1977) 1311.
\bibitem{13} J.L.Lebowitz, Phys. Rev., 133 (1964) 895.
\bibitem{14} M.V.Tokarchuk, A.E.Kobryn, Y.A.Humenyuk, Lviv, Preprint ICMP-99-08U.
        It is available also on http://www.ICMP.Lviv.UA/ICMP/preprints/PS/9908Ups.gz
\bibitem{15} N.B.Vargaftik, Handbook of Physical Properties of Liquids and Gases,
        Begell House Inc., NY, 1996.
\bibitem{16} A.E.Kobryn, I.P.Omelyan, M.V.Tokarchuk, J. Stat. Phys., 92
        (1998) 973.
\end{thebibliography}
\end{document}